\documentclass[12pt]{iopart}
\usepackage[dvips]{graphicx}

\begin{document}

%\draft
%\preprint{}
\title{Formation of clusters in the $2D$ $t-J$ model:
The mechanism for phase separation.}

\author{A. Fledderjohann\dag\, A. Langari\ddag
\footnote[3]{To whom correspondence should be addressed
(langari@mpipks-dresden.mpg.de)}
and K.-H. M\"utter\dag\
}

\address{\dag\ Physics Department, University of Wuppertal, 42097
Wuppertal, Germany}

\address{\ddag\ Department of Physics, Sharif University of Technology, 
Tehran 11365-9161, Iran
%\ddag\ Institute for Studies in Theoretical Physics and Mathematics (IPM), Tehran 19395-5531, Iran
}

\date{\today}

%\maketitle

\begin{abstract}
\leftskip 2cm
\rightskip 2cm
%%%%%%%%%%%%%%%%%%%%%%%%%%%%%%%%%%%%%%%%%%%%%%%%%%%%%%%%%%%%%%%%%%%%%%%%%%%%
The emergence of phase separation is investigated in the framework of a $2D$
$t-J$ model by means of a variational product ansatz, which covers the infinite
lattice by two types of $L\times L$ clusters. Clusters of the first type are
completely occupied with electrons, i.e. they carry maximal charge $Q_e=L^2$ and
total spin 0, and thereby form the antiferromagnetic background. Holes occur
in the second type of clusters -- called ``hole clusters''. They carry a 
charge $Q_h<L^2$. The charge $Q_h$ and the number $N(Q_h)$ of hole clusters is fixed
by minimizing the total energy at given hole density and spin exchange
coupling $\alpha=J/t$. For $\alpha$ not too small ($\alpha>0.5$) it turns out
that hole clusters are occupied with an even number $Q_h<L^2$ of electrons and
carry a total spin 0. For increasing $\alpha$ the charge $Q_h(\alpha)$ of the hole
clusters decreases. %For $\alpha>3.5$ it turns out to be zero.
%Moreover, the interaction energy between neighbouring clusters %plaquettes ($L=2$) favours 
%favours the condensation of clusters with equal charge,
%which implies phase separation. 
%abdollah-a1
%It's
Some points on the
 boundary curve can be extracted from $Q_h(\alpha)$.
%product ansatz for the ground state with $L\times L$
%clusters ($L=2$, $L=4$). The minimum of the ground state energy is 
%characterized by two cluster charges $Q_1,Q_2$, which depend on the charge
%density $\rho=Q/N$ and the spin exchange coupling $\alpha=J/t$. Signatures
%for phase separation are found in those domains of the $\rho-\alpha$ plane,
%where the two cluster charges $Q_1,Q_2$ are well ``separated'' [($Q_1,Q_2$)=
%(0,4), (2,4) for $L=2$; ($Q_1,Q_2$)= (0,16), (2,16), (6,16), (10,16), (14,16)
%for $L=4$]. Moreover, the interaction energy between neighbouring plaquettes
%clusters ($L=2$) favours the condensation of clusters with equal charge.
%%%%%%%%%%%%%%%%%%%%%%%%%%%%%%%%%%%%%%%%%%%%%%%%%%%%%%%%%%%%%%%%%%%%%%%%%%%%
\end{abstract}

\pacs{71.10.Fd,71.27.+a,75.10.-b, 75.10.Jm}
% Uncomment for Submitted to journal title message
\submitto{Journal of Physics: Condensed Matter}

\maketitle

%%%%%%%%%%%%%%%%%%%%%%%%%%%%%%%%%%%%%%%%%%%%%%%%%%%%%%%%%%%%%%%%%%%%%%%%%%%%
\section{Introduction\label{sec1}}
%\section{Introduction: Motivation and questions of interest}
%\setcounter{equation}{0}

Holes play a fundamental role in our understanding of high $T_c$
superconductivity.\cite{bednorz86} The parent materials like
$La_2CuO_4$ are insulators with antiferromagnetic order and
doping with holes (missing electrons) opens the superconducting
phase. \cite{anderson87}

Experimental evidence has been found in $La_2CuO_{4+\delta}$
for phase separation.\cite{jorgensen88,hammel90,hammel91,chou96}
This means that the holes in the $CuO_2$ planes are not
distributed uniformly but concentrate in ``hole rich''
domaines. 
%abdollah-new
The compound phase separates for $0.01\leq \delta \leq 0.06$ below $T_{ps} \sim 300 K$
into the nearly  stoichiometric antiferromagnetic  $La_2CuO_{4+\delta_1}$ with
$\delta_1 < 0.02$ and N\'eel temperature ($T_N \simeq 250 K$), and a metallic superconducting
oxygen-rich phase $La_2CuO_{4+\delta_2}$ with $\delta_2 \approx 0.02$ and $T_c \simeq 34 K$.
The other evidence is related to the $Sr$ doped compound $La_{2-x}Sr_xCuO_{4+\gamma}$.
This compound phase separates for $x \leq 0.03$ into the superconducting $La_{2-x}Sr_xCuO_{4+\gamma_1}$
($\gamma_1\approx 0.08$) and the nonsuperconducting $La_{2-x}Sr_xCuO_{4+\gamma_2}$
($\gamma_2\approx 0.00$) phases. \cite{cho93}
%end-abdollah-new

Intensive studies have been performed to understand this phenomenon
in models for strongly correlated electrons like the Hubbard --
and $t-J$ model.
In the $2D$ $t-J$ model various attempts have been made
to exploit the phase diagram in the plane spanned by the
charge density $\rho=Q_{\mbox{tot}}/N$ and the spin coupling
$\alpha=J/t$ ($N$, $Q_{\mbox{tot}}$ denote the total numbers
of sites and electrons, respectively).

Phase separation occurs, if both the charge density $\rho$ and
the spin coupling $\alpha$, are large enough
\begin{eqnarray}
1\geq\rho\geq\rho_1 & , & \alpha>\alpha_p(\rho_1)\,.\label{large_rho}
\end{eqnarray}
In this regime, the ground state can be represented by
a product ansatz
\begin{equation}
\psi(\rho,N) =\psi_e\Big(N_e(\rho)\Big)\psi_h\Big(
N_{h}(\rho,\rho_1)\Big)
\label{psiprod}
\end{equation}
of two clusters with site numbers $N_e(\rho)$ and $N_{h}(\rho,\rho_1)$,
which cover the whole lattice:
\begin{eqnarray}
N_e(\rho)+N_{h}(\rho,\rho_1) & = & N\,.\label{clusters}
\end{eqnarray}
The $N_e(\rho)$ sites in the ``electron'' cluster are all occupied with
electrons. The corresponding ground state $\psi_e(N_e(\rho))$
is just the ground state of the $2D$ Heisenberg model with $N_e(\rho)$ sites.
Holes occur in the ``hole'' cluster with $N_h(\rho,\rho_1)$ sites.
The corresponding ground state $\psi_h(N_h(\rho,\rho_1))$ is given by the
$t-J$ model with $N_h(\rho,\rho_1)$ sites and $Q_h=N_h(\rho,\rho_1)\rho_1$
electrons. The numbers $N_e(\rho)$ and $N_h(\rho,\rho_1)$ for the cluster
sites are fixed by the total number of sites (\ref{clusters}) and the
total number of electrons
%
%
%In the second cluster, the charge density is fixed to be $\rho_1$.
%The corresponding wave function $\psi_{\rho_1}(
%N_{\rho_1}(\rho))$ is given by the $t-J$ model on a cluster with
%$N_{\rho_1}(\rho)$ sites and $Q_1=N_{\rho_1}(\rho)\rho_1$ electrons.
%The numbers $N_1(\rho)$ and $N_{\rho_1}(\rho)$ for the cluster sites
%are fixed by the total number of sites (\ref{clusters}) and the total
%number of electrons: % (\ref{clusters})
%
\begin{eqnarray}
N_e(\rho)+\rho_1N_{h}(\rho,\rho_1) & = & Q_{\mbox{tot}}\,.\label{clusters2} 
\end{eqnarray} 
From (\ref{psiprod})-(\ref{clusters2}) one derives that the
ground state energy per site $\varepsilon(\rho,\alpha)$:
\begin{eqnarray}
\varepsilon(\rho,\alpha) & = & \frac{1}{1-\rho_1}\left[
\varepsilon(\rho_1,\alpha)(1-\rho)+\varepsilon(1,\alpha)(\rho-\rho_1)\right]
\label{e_lin_rho}
\end{eqnarray}
is linear in $\rho$. This holds for (\ref{large_rho}), i.e.
in the region with phase separation.

The focus %of 
%abdollah-a1
%interest
%some studies
 is the boundary curve $\alpha_p(\rho_1)$
for phase separation.
Two controversary points of view can be found in the literature:
Emery et al.\cite{emery90} suggested that in the $2D$ $t-J$
model the phase separation curve $\alpha_p(\rho_1)$ starts at 
$\alpha_p(\rho_1=1)=0$, which means that phase separation exists
already for low doping $(1-\rho_1)\simeq 0$ and small values of $\alpha$,
as observed experimentally.\cite{jorgensen88}
Their point of view is supported by Hellberg and Manousakis
\cite{hellberg95,hellberg97,hellberg99,hellberg00}.

On the other hand Putikka et al.\cite{putikka92,putikka94} concluded
from a high temperature expansion, that phase separation
at low doping only emerges for larger $\alpha$ values
$\alpha>\alpha_p(\rho_1=1)\simeq 1.2$, which would exclude
the $\alpha$ regime realized in the experiments. 
This point of view is supported by DMRG calculationss on
ladders (Rommer, White, Scalapino\cite{rommer00}), 
variational wave functions of
the Luttinger-Jastrow-Gutzwiller type (Valenti, Gros\cite{valenti92}),
Lanczos calculations (Dagotto\cite{dagotto92,dagotto98}). 

%abdollah-a1
Finally, Green's function Monte-Carlo simulations performed by Calandra et al. (CBS)\cite{calandra98}
led to a value $\alpha_p(\rho_1=1)\simeq 0.5$.
%and their modifications (Fehske\cite{fehske91}).

%In this paper we would like to propose a new method
%which extracts $\alpha_p(\rho_1)$ from the ground state energies
%$E^{(L)}(Q,\alpha=\alpha(\rho_1=Q_1/N))$
%on $L\times L=N$ clusters with charges $Q=L^2, Q=Q_1, Q=Q_1-L$.
%These energies can be computed analytically for plaquette
%clusters ($L=2$) as will be discussed in Section \ref{sec2}.
%Phase separation can be observed in these small clusters on a
%``microscopic scale''.
%The interaction energies $W(Q,Q')$ between neighbouring
%plaquette clusters are discussed in Section \ref{sec3}.
%They favour clustering of plaquettes with equal charge,
%which can be interpreted as phase separation on a
%``macroscopic scale''.

%abdollah-a1
%However
In this paper, we would like to study the mechanism of phase separation by
means of a product ansatz with $L\times L$ clusters of charge $Q$
which cover the infinite lattice.
Our method is in the spirit of the coupled cluster method (CCM)
designed almost 50 years ago \cite{coester58} as an 
approximation scheme for quantum many body problems. In order to
handle the interaction between clusters, perturbative \cite{gros93}
and variational \cite{potthoff03} methods have been
implemented. 
In our study of the phase diagram in the ($\rho=Q/N,\alpha=J/t$)
plane, we proceed as follows: We start from the ground states
$\psi^{(L)}(\rho_1,\alpha)$ with energies $E^{(L)}(Q_1,\alpha)$
on $L\times L$ clusters with charge $Q_1=\rho_1L^2$. These ground states
can be computed analytically for plaquette clusters ($L=2$).
In Section \ref{sec2} we will discuss the product ansatz with
plaquette clusters, which minimizes the total energy at fixed
charge density $\rho=Q/N$. Phase separation can be observed
on the small plaquette cluster on a ``microscopic scale''.
%The interaction energies $W(Q,Q')$ between neighbouring plaquette
%clusters with total charges $Q,Q'$ are discussed in Section \ref{sec3}. 
%For $\alpha$ values not too small, they favour clustering
%of plaquettes with equal charge $Q=Q'$, which can be interpreted as phase
%separation on a ``macroscopic scale''.

In Section \ref{sec4} we extend our considerations to 
clusters of size $L\times L$. Results for the phase diagram
obtained from a numerical calculation
of ground state energies on a $4\times 4$ cluster are shown. 
%abdollah-a1
We finally summerize and present the discussion in Section \ref{sec6}.

%presented.
%Section \ref{sec5} presents results on hole--hole correlators
%obtained from our cluster product ansatz.

%%%%%%%%%%%%%%%%%%%%%%%%%%%%%%%%%%%%%%%%%%%%%%%%%%%%%%%%%%%%%%%%%%%%%%%%%%%%%%
\section{Plaquette cluster in the $t-J$ model: Phase
separation ``in statu nascendi''\label{sec2}}

\begin{figure}[ht!]
\centerline{\hspace{0.0cm}\includegraphics[width=6.0cm,angle=0]{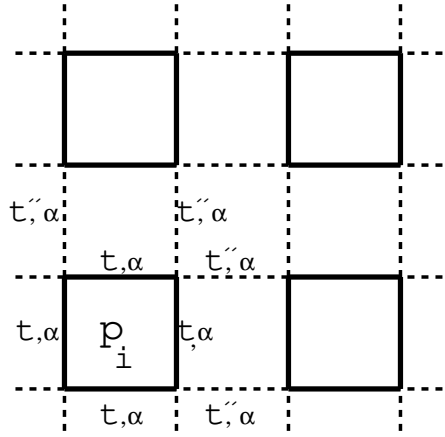}
\hspace{-0.0cm}}
\caption{$2D$ lattice structure with underlying $2\times 2$ plaquette
structure.}
\label{fig1}
\end{figure}

We start from the $t-J$ model in two dimensions which is built up
as a lattice of $L\times L$ clusters (cf. Fig. \ref{fig1} for the
example of $L=2$):
\begin{eqnarray}
H & = & t\sum_i h_{p_i}(\alpha=J/t)+t''\sum_{\langle i,j\rangle}
h_{p_i,p_j}(\alpha)\label{H_t_tpp}
\end{eqnarray}
with $h_{p_i}$ and $h_{p_i,p_j}$ containing all nearest neighbour
bonds $\langle k,l\rangle$ in a cluster $p_i$ or between neighbouring
clusters ($p_i$, $p_j$), respectively.
In general, each nearest neighbour bond $\langle k,l\rangle$ of the lattice either belongs
to a cluster (parameters $t,J=\alpha t$) or connects neighbouring
clusters ($t'',J''=\alpha t''$) and contributes with $th^{(k,l)}(\alpha)$
or $t''h^{(k,l)}(\alpha)$
to the respective intra- or inter-cluster part of the Hamiltonian (\ref{H_t_tpp}):
%(i.e. $th^{(k,l)}(\alpha)$ or $t''h^{(k,l)}(\alpha)$):
%-- e.g. for $\langle k,l\rangle$ as 
%nearest neighbour bond
%in a cluster $p_i$ the corresponding contribution $h_{p_i}^{(j,k)}(\alpha)$ 
%to $h_{p_i}(\alpha)$ reads
\begin{equation}
h^{(k,l)}(\alpha)= 
{\cal P}\Big[-\sum_{\sigma}
\left(
c_{k,\sigma}^+
c_{l,\sigma} + \mbox{h.c.} \right)
+\alpha \Big({\bf S}_{k}\cdot {\bf S}_{l}
-\frac{1}{4}n_{k}n_{l}\Big)\Big]{\cal P} 
\label{element_h_plaq}
\end{equation}
%\begin{eqnarray}
%H & = & t\sum_{i, j\in p_i} h_{i, j}(\alpha=J/t)+t''\sum_{i\in p_i ,j\in p_j}
%h_{i, j}(\alpha).%\nonumber\\
%\end{eqnarray}
%Each nearest neighbour bond in Fig. \ref{fig1}
%either belongs to a cluster ($p_i$; parameters $t,J=\alpha t$) or connects neighbouring
%clusters ($p_i, p_j$; $t'',J''=\alpha t''$).
%The Hamiltonian of each bond ($h_{j, k}(\alpha)$) reads
%\begin{eqnarray}
%h_{j, k}(\alpha) & = &
%{\cal P}\Big[-\sum_{\sigma}
%\left(
%c_{j,\sigma}^+
%c_{k,\sigma} + \mbox{h.c.} \right)\nonumber\\
% &  &
%\hspace{0.5cm}+\alpha \Big({\bf S}_{j}{\bf S}_{k}
%-\frac{1}{4}n_{j}n_{k}\Big)\Big]{\cal P},\nonumber\\ \label{element_h_plaq}
%\end{eqnarray}
where
\begin{eqnarray}
n_{l} & = & \sum_{\sigma}n_{l,\sigma}=\sum_{\sigma}c_{l,\sigma}^+
c_{l,\sigma}
\end{eqnarray}
and ${\cal P}$ projects onto the subspace where 
the occupation numbers $n_{l}$ are restricted to zero or one.
%abdollah-2 The following line has been added:
The electron spin operator is represented by ${\bf S}_{k}$ at the k-th site.

%The intercluster terms are built correspondingly 
%with hopping parameter $t''$. 
%except the double-primed
%The latter are meant to emphasize the difference
%to the zeroth order situation
%of uncoupled clustersr. 
%Finally, we are aiming to obtain lowest order
%perturbation results for the %homogeneous 
%isotropic $2D$ $t-J$ lattice (i.e.
%$t''=t\equiv 1$). Before, we will discuss the
%physical content obtainable from single clusters ($t''=0$).
%The question of choosing appropriate boundary conditions for the
%$L\times L$ clusters is of particular interest (see e.g.
%[\onlinecite{note1}] ***(??)***). \ldots

%\vspace{2.0cm}
%(which are --due to the projection operator $\cal{P}$-- restricted 
%to values zero or one, i.e. no double occupancy of lattice
%sites by electrons of opposite spin)
%as shown in Fig. \ref{fig1}. 

%In particular for the isotropic case ($t''=t$, $\alpha''=\alpha$)
%we write
%\begin{eqnarray}
%H & = & t\left(H_t+\alpha H_J\right);\hspace{1.0cm} \alpha=\frac{J}{t}\,.
%\end{eqnarray}

The interaction between $L=2$ clusters is mediated by
the two dashed links for each spatial dimension. It is treated in the following by a
perturbation expansion in the hopping parameter $t''$, which
allows the hopping of electrons between neighbouring plaquettes.
%(****** remark on $\alpha''$?? ******)
%%%%%%%%%%%%%%%%%%%%%%%%%%%%%%%%%%%%%%%%%%%%%%%%%%%%%%%%%%%%%%%%%%%%%%%%
%%%%%%%%%%%%%%%%%%%%%%%%%%%%%%%%%%%%%%%%%%%%%%%%%%%%%%%%%%%%%%%%%%%%%%%%

To zeroth order in $t''$, the eigenstates of the $2D$ $t-J$ 
model decay into a product
\begin{eqnarray}
%|Q_1\ldots Q_{N^{(p)}}\rangle & = & \prod_p |Q_p\rangle
 & & \prod_p |Q_p\rangle\label{prod_ansatz}
\end{eqnarray}
of plaquette eigenstates $|Q_p\rangle$ with charge $Q_p$.
The ground state is characterized by a minimum of the energy:
\begin{eqnarray}
E_0=\sum_p E^{(p)}(Q_p) & = & \sum_{Q=0}^4 E^{(p)}(Q)
N(Q)\label{e0_I}
\end{eqnarray}
where $N(Q)$ is the number of plaquettes with charge $Q=0,1,2,3,4$.
$E^{(p)}(Q)$ is the ground state energy of a plaquette with charge
$Q$.

The numbers $N(Q)$ are constrained by the total number of plaquettes ($N/4$)
and the total charge ($Q_{\mbox{tot}}$):
%\begin{itemize}
%\item
%the total number of plaquettes $N^{(p)}$
\begin{eqnarray}
\sum_{Q=0}^4 N(Q)=N^{(p)}=\frac{N}{4}%\label{e0_II}
 & \quad,\quad & \sum_{Q=0}^4 QN(Q)=Q_{\mbox{tot}}\label{e0_II}
\end{eqnarray}

%\item
%the total charge
%\begin{eqnarray}
%\sum_{Q=0}^4 QN(Q) & = & Q_{\mbox{tot}}\label{e0_III}
%\end{eqnarray}
%\end{itemize}
The plaquette ground state energies $E^{(p)}(Q)$ have been computed in
Appendix A of Ref. \cite{fl04}. % for the general case with
%different couplings $t,\alpha$ %and $t',\alpha'$ 
%in the vertical
%and horizontal direction. 
The there obtained formulas simplify for the
symmetric case %$t=t',\,\alpha=\alpha'$ 
we are considering here.

%hier_8
In the charge sectors $Q=0,1,2,4$
there is no level crossing and the plaquette
ground state energy for all $\alpha>0$ reads
\begin{eqnarray}
E^{(p)}(0) & = & 0\hspace{4.0cm}S=0\label{eQ0_1}\\
E^{(p)}(1) & = & -2t\hspace{3.4cm}S=1/2\label{eQ1_1}\\
E^{(p)}(2) & = & -\frac{t}{2}\Big(\alpha+\sqrt{\alpha^2+32}\Big)\hspace{0.55cm}S=0\label{eQ2_1}\\
E^{(p)}(4) & = & -3t\alpha\hspace{3.15cm}S=0\label{eQ4_1}
\end{eqnarray}
Note that the total spin ${\bf S}^{(p)}$ of the plaquette
electrons commutes with the plaquette Hamiltonian and the
eigenvalues of ${{\bf S}^{(p)}}^2=S(S+1)$ can be used to
characterize the plaquette ground state.
This plays an important role in the sectors with
$Q=3$ where the ground state changes with $\alpha$:
\begin{eqnarray}
E_{I}^{(p)}(3,\alpha) & = & -2t,\hspace{31mm}S=3/2 \hspace{3mm}\mbox{for}
 \hspace{3mm}0\leq\alpha\leq\alpha_m\nonumber\\
[4pt]
E_{II}^{(p)}(3,\alpha) & = &  -t\Big(\alpha+\sqrt{\frac{\alpha^2}{4}
+3}\Big),\hspace{5mm}S=1/2\hspace{5mm}\mbox{for}
\hspace{5mm}\alpha_m\leq\alpha<2\nonumber\\[4pt]
E_{III}^{(p)}(3,\alpha) & = & -t\Big(\frac{3\alpha}{2}+1\Big),
\hspace{17mm}S=1/2\hspace{5mm}\mbox{for}
\hspace{5mm}\alpha>2. \label{eQ3}
\end{eqnarray}
with $\alpha_m=\frac{2}{3}(4-\sqrt{13})=0.262..$.

%It is remarkable to note,
Note, that the Nagaoka ferromagnetic state\cite{nagaoka66} --
with maximal spin $S=Q/2$ -- is already visible on a 4-site plaquette 
with one hole, i.e. $Q=3$, if $\alpha$ is small enough $\alpha<
\alpha_m=0.262$. Eigenstates with maximal plaquette spin $S=Q/2$
exist in all charge channels; the corresponding energy eigenvalues
$E(Q,S=Q/2)$ do not depend on the spin exchange coupling $\alpha$. However,
these eigenstates are not ground states except for the one hole case.

Let us next look for the minimum of the total energy (\ref{e0_I}),
which depends on the relative magnitude of the plaquette energies.
\begin{itemize}
\item[a)]
In the regime, where the inequality
\begin{eqnarray}%\label{ineq_4233}
E^{(p)}(4)+E^{(p)}(2) & < & 2E^{(p)}(3)
%\nonumber\\ 
\label{ineq_4233}
\end{eqnarray}
holds, two plaquettes with charge $Q=3$ are %eliminated and
substituted by two plaquettes with charges $Q=2$
and $Q=4$, respectively.
Insertion of the ground state energies [(\ref{eQ2_1}), (\ref{eQ4_1}),
(\ref{eQ3})]
%for $Q=2$, (\ref{eQ4_1}) for $Q=4$ and
%(\ref{eQ3})(I,II,III) for $Q=3$
yields the validity of (\ref{ineq_4233}) for
\begin{eqnarray}\label{al_ineq_4233}
\alpha & > & \alpha(3)=\frac{2}{\sqrt{21}}\simeq 0.436 \,.
\label{regime_a}
\end{eqnarray}

\item[b)]
The inequalitiy
\begin{eqnarray}
E^{(p)}(0)+E^{(p)}(4) & < & 2E^{(p)}(2)
\label{ineq_0422}
\end{eqnarray} 
holds for
\begin{eqnarray}
\alpha & \geq & \alpha(2)=4\sqrt{2/3}\simeq 3.266
\label{regime_b}
\end{eqnarray}
and allows the substitution of the charge $Q=2$ plaquettes.

\item[c)]
The inequality
\begin{eqnarray}
E^{(p)}(0)+E^{(p)}(2) & < & 2E^{(p)}(1)%\nonumber
\label{ineq_0211}
\end{eqnarray}    
holds for 
\begin{eqnarray}
\alpha & > & 2
\label{regime_c}
\end{eqnarray}
and allows to substitute $Q=1$ plaquettes.

\end{itemize}

\begin{figure}[ht!]
\centerline{\hspace{0.0cm}\includegraphics[width=15.0cm,angle=0]{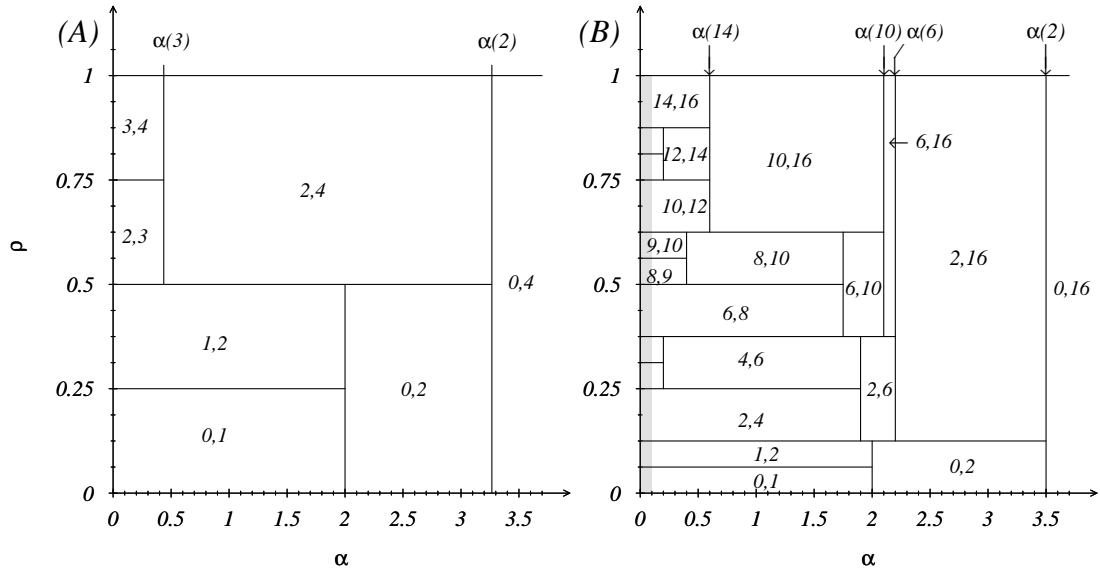}
\hspace{0.0cm}}
\caption{Schematic phase diagram obtained from $2\times 2$ plaquettes (A)
%The charges $Q_1,Q_2$ that constitute the respective ground states
%in the $\rho-\alpha$-plane are given and the boundaries %$\alpha_{1,2}$
%of the $(2,4)$ regime are: $\alpha(2)=2/\sqrt{21}$, $\alpha(0)=4\sqrt{2/3}$;
and $4\times 4$ plaquettes (B).
The charges $Q_1,Q_2$ that constitute the respective ground states
in the $\rho-\alpha$-planes and the boundaries $\alpha(2j)$ are given in the text.}
\label{fig2}
\end{figure}

Fig. \ref{fig2}(A) shows a schematic view of the phase diagram in
the $\rho -\alpha$ plane with the pairs of plaquette charges,
which fix the ground state energy per site according to eqs.
(\ref{e0_I})-(\ref{e0_II}).

%%%%%%%%%%%%%%%%%%%%%%%%%%%%%%%%%%%%%%%%%%%%%%%%%%%%%%%%%%%%%%%%%%%%%%%%
%%%%%%%%%%%%%%%%%%%%%%%%%%%%%%%%%%%%%%%%%%%%%%%%%%%%%%%%%%%%%%%%%%%%%%%%

In each of the domains with plaquette pairs $Q_1,Q_2$ 
\begin{eqnarray}
Q_1<Q_2 & \quad\quad & \frac{Q_1}{N_c}<\rho<\frac{Q_2}{N_c}
\end{eqnarray}
%abdollah-a1
where the  plaquette sites $N_c=4$,  the ground state energy per site:
\begin{equation}
\varepsilon(\rho,\alpha)=\frac{1}{Q_2-Q_1}\Big\{
E^{(p)}(Q_1)\Big(\frac{Q_2}{N_c}-\rho\Big)+
E^{(p)}(Q_2)\Big(\rho-\frac{Q_1}{N_c}\Big)\Big\}
\label{ge_Q1Q2}
\end{equation}
is linear in $\rho$ and the chemical potential
$\mu(\rho,\alpha)=d\varepsilon/d\rho$
%\begin{eqnarray}
%\mu(\rho,\alpha) & = & \frac{d\varepsilon}{d\rho}
%\label{chemical_pot}
%\end{eqnarray}
is constant with respect to $\rho$.
%\begin{eqnarray}
%\mu(\rho,\alpha) & = & \frac{1}{Q_2-Q_1}\Big\{
%E^{(p)}(Q_2,\alpha)-E^{(p)}(Q_1,\alpha)\Big\}\,.
%\end{eqnarray}
Note that two domains ($Q_1,Q_2$) and ($Q_2,Q_3$) with a common
plaquette charge $Q_2$ have also a common horizontal boundary at charge
density $\rho_2=Q_2/4$.
At this boundary the chemical potential $\mu(\rho,\alpha)$
as function of $\rho$ is discontinuous with a jump
\begin{eqnarray}
\Delta(Q_2,\alpha) & = & \mu(\rho+0,\alpha)-\mu(\rho-0,\alpha)\nonumber\\
 & = & \frac{E^{(p)}(Q_3)}{Q_3-Q_2}+\frac{E^{(p)}(Q_1)}{Q_2-Q_1}-
\frac{Q_3-Q_1}{(Q_3-Q_2)(Q_2-Q_1)}E^{(p)}(Q_2)
\end{eqnarray}
which vanishes for a specific value of $\alpha=\alpha(Q_2)$.
%\begin{eqnarray}
%\Delta(Q_2,\alpha=\alpha(Q_2)) & = & 0\,.
%\end{eqnarray}
They define the vertical lines in the phase diagram given by
(\ref{regime_a}), (\ref{regime_b}) and (\ref{regime_c}).
This type of phase diagram has been introduced first by Kagan et al.
\cite{kagan99} 
for the three leg ladder in the rung cluster approximation.
Note, that the constraints for ($N^{(p)}$, $Q_{\mbox{tot}}$)(\ref{e0_II})
%,(\ref{e0_III})
are taken into account explicitely on both sides of the inequalities
(\ref{ineq_4233}), (\ref{ineq_0422}), (\ref{ineq_0211}).
This procedure can be considered as an alternative to the usual
grand canonical one, where the charge conservation constraint (\ref{e0_II})
is eliminated by a reservoir with a fixed chemical potential.
\section{Product ansatz for the ground state with large clusters\label{sec4}}

The considerations which led to the phase diagram in Fig. \ref{fig2}(A) based
on plaquette clusters can easily be extended to larger $L\times L=N_c$
clusters with $N_c$ sites and charges $Q=N_c,\ldots,0$. We start with
the ground state energies $E(Q,N_c,\alpha)$ and their dependence
on $\alpha$. 
The generalization of (\ref{ge_Q1Q2}) to a cluster
with $N_c$ sites and charges $Q_1<Q_2$ is straightforward
%abdollah-a1
by replacing $E^{(p)}(Q_1) \rightarrow E(Q_1,\alpha,N_c)$.
%\begin{eqnarray}
%\varepsilon(\rho,\alpha) & = & \frac{1}{Q_2-Q_1}\Big\{
%E(Q_1,\alpha,N_c)\Big(\frac{Q_2}{N_c}-\rho\Big)+
%E(Q_2,\alpha,N_c)\Big(\rho-\frac{Q_1}{N_c}\Big)\Big\}.
%\label{ge_Q1Q2_II}
%\end{eqnarray}
Again the ground state energy per site is linear in $\rho$ and the
chemical potential is constant:
\begin{eqnarray}
\mu(\rho,\alpha) & = & \frac{1}{Q_2-Q_1}\Big\{
E(Q_2,\alpha,N_c)-E(Q_1,\alpha,N_c)\Big\}\,.
\label{mu_rho}
\end{eqnarray}
Two domains ($Q_1,Q_2$) ($Q_2,Q_3$) with a common cluster charge $Q_2$
have a common boundary in the phase diagram at $\rho_2=Q_2/N_c$. At this
boundary the chemical potential is discontinuous with a jump
\begin{eqnarray}
\Delta(Q_2,\alpha) & = & \mu(\rho_2+0,\alpha)-\mu(\rho_2-0,\alpha)
\nonumber\\
 & = & \frac{E(Q_3)-E(Q_2)}{Q_3-Q_2}-\frac{E(Q_2)-E(Q_1)}{Q_2-Q_1}
\label{jump}
\end{eqnarray}
which vanishes for a specific value of $\alpha=\alpha(Q_2)$.
%abdollah-a1
%\begin{eqnarray}
%\Delta(Q_2,\alpha(Q_2)) & =  & 0\,.
%\label{Delta_Q2_0}
%\end{eqnarray}
The inequality
\begin{eqnarray}
\Delta(Q_2,\alpha) & <  & 0\quad\mbox{for\quad} \alpha>\alpha(Q_2)
\end{eqnarray}
means that the clusters with charge $Q_2$ and ground state energy
$E(Q_2,\alpha,N_c)$ can be substituted by clusters with charges
$Q_3$ and $Q_1$ and energies $E(Q_3,\alpha,N_c)$, $E(Q_1,\alpha,N_c)$.
Here the two domains ($Q_1,Q_2$) ($Q_2,Q_3$) merge 
%together:
%abdollah-a1
for $ \alpha>\alpha(Q_2)$
%\begin{eqnarray}
%(Q_1,Q_2), (Q_2,Q_3) & \rightarrow & (Q_1,Q_3)\quad
%\mbox{for\quad} \alpha>\alpha(Q_2)
%\end{eqnarray}
and the ground state is given by a cluster product ansatz with
charges ($Q_1,Q_3$).

Let us first consider the case [case ($A$)]
\begin{eqnarray}
Q_1=Q_2-1,\,\,Q_3=Q_2+1, \,\, & & Q_2\quad\mbox{odd},\,\,\alpha<\alpha(Q_2)\,.
\end{eqnarray}
Here the ground state is built up from clusters with charges:
\begin{eqnarray}
(Q_2-1,\,Q_2) & \hspace{0.5cm}\mbox{for}\hspace{0.5cm} & 
\frac{Q_2-1}{N_c}\leq\rho\leq\frac{Q_2}{N_c}
\label{domain1}\\
(Q_2,\,Q_2+1) & \hspace{0.5cm}\mbox{for}\hspace{0.5cm} & \frac{Q_2}{N_c}\leq\rho\leq\frac{Q_2+1}{N_c}\,.
\label{domain2}
\end{eqnarray}
The vanishing of the jump (\ref{jump}) in the chemical potential
\begin{eqnarray}
\Delta(Q_2,\alpha(Q_2)) & = & E(Q_2+1)+E(Q_2-1)-2E(Q_2)=0
\label{Delta_0}
\end{eqnarray}
defines the boundary $\alpha=\alpha(Q_2)$, where the two domains
(\ref{domain1}) and (\ref{domain2}) merge together:
\begin{eqnarray}
(Q_2-1,\,\,Q_2+1)\quad\mbox{for\quad}\frac{Q_2-1}{N_c}\leq\rho\leq\frac{Q_2+1}{N_c}\hspace{5mm}
\mbox{and}\,\,\alpha>\alpha(Q_2). 
\end{eqnarray}
The numerical evaluation of (\ref{Delta_0}) from the ground state energies
on a $4\times 4=16$ cluster -- with periodic boundary conditions -- yields
the couplings $\alpha(Q_2)$ listed in Table \ref{table_n1} (col. 1,2):

\begin{table}[t]
\begin{center}
\caption{Boundary couplings $\alpha(Q_2)$ [from left to right cases ($A$), ($B$), ($C$)] where
the jump (\ref{jump}) in the chemical potential vanishes for cases $(Q_1,Q_2,Q_3)$.}
\begin{tabular}{c|c||c|c||c|c}
 $(Q_1,Q_2,Q_3)$ &  $\alpha(Q_2)$  &  $(Q_1,Q_2,Q_3)$ &  $\alpha(Q_2)$  &
 $(Q_1,Q_2,Q_3)$ &  $\alpha(Q_2)$ \\ \hline
 $(14,15,16)$ & $<0.1$ & $(12,14,16)$ & $0.6$ & $(6,10,16)$ & $2.1..$\\
 $(12,13,14)$ & \hspace{0.4cm}$0.2$ & $(10,12,14)$ & $0.6$ & $(2,6,16)$ & $2.2..$\\
 $(10,11,12)$ & $<0.1$ & $(6,8,10)$ & $1.74$ & $(0,2,16)$ & $3.5..$\\
 $(8,9,10)$ & \hspace{0.4cm}$0.4$ & $(2,4,6)$ & $1.9$ &  & \\
 $(6,7,8)$ & $<0.1$ & & &  & \\
 $(4,5,6)$ & \hspace{0.4cm}$0.2$ & & &  & \\
 $(2,3,4)$ & $<0.1$ & & &  & \\
 $(0,1,2)$ & \hspace{0.4cm}$2.0$ & & &  & 
\end{tabular}
\label{table_n1}
\end{center}
\end{table}

%\begin{table}[t]
%\begin{center}
%\caption{Boundary couplings $\alpha(Q_2)$ for odd $Q_2$ where
% (\ref{jump})  vanishes, for $(Q_1,Q_2,Q_3)=(Q_2-1,Q_2,Q_2+1)$.}
%\begin{tabular}{c|c|c|c|c|c|c|c|c}
% $Q_2$ &  $15$ &
% $13$
% &  $11$ & 9 & 7 & 5 & 3 & 1
%\\ \hline
% $\alpha(Q_2)$ &  $<0.1$ &
% $0.2$
% & $<0.1$ & 0.4 & $<0.1$ & 0.2 & $<0.1$ & 2.0
%\end{tabular}
%\label{table1}
%\end{center}
%\end{table}

In the next step, we consider the cases [case ($B$)]:
\begin{eqnarray}
Q_1=Q_2-2,\,\,Q_3=Q_2+2,\hspace{5mm}
Q_2=14,12,8,4, \hspace{5mm}\alpha<\alpha(Q_2)\,.
\end{eqnarray}
Here the ground state is built up from clusters with charges
\begin{eqnarray}
(Q_2-2,\,Q_2) & \hspace{0.5cm}\mbox{for}\hspace{0.5cm} & \frac{Q_2-2}{N_c}\leq\rho\leq\frac{Q_2}{N_c}
\label{domain3}\\
(Q_2,\,Q_2+2) & \hspace{0.5cm}\mbox{for}\hspace{0.5cm} & \quad\frac{Q_2}{N_c}\leq\rho\leq\frac{Q_2+2}{N_c}\,.
\label{domain4}
\end{eqnarray}
The vanishing of the jump (\ref{jump}) in the chemical potential
\begin{eqnarray}
\Delta(Q_2,\alpha(Q_2)) & = & \frac{1}{2}\left[E(Q_2+2)+E(Q_2-2)-2E(Q_2)
\right]= 0
\label{Delta_02}
\end{eqnarray}
defines again the boundary $\alpha=\alpha(Q_2)$, where the two domains
(\ref{domain3}) and (\ref{domain4}) merge together:
\begin{equation}
(Q_2-2,\,\,Q_2+2)\quad\mbox{for\quad}\frac{Q_2-2}{N_c}\leq\rho\leq\frac{Q_2+2}{N_c}\hspace{5mm}
\mbox{and} \hspace{5mm}\alpha>\alpha(Q_2). 
\end{equation}
The numerical evaluation of (\ref{Delta_02}) from the ground state energies
on a $4\times 4=16$ cluster -- with periodic boundary conditions -- yields
the couplings $\alpha(Q_2)$ listed in Table \ref{table_n1} (col. 3,4):

%\begin{table}[t]
%\begin{center}
%\caption{Boundary couplings $\alpha(Q_2)$ for even $Q_2=14,12,8,4$ where 
% (\ref{jump}) vanishes, for $(Q_1,Q_2,Q_3)=(Q_2-2,Q_2,Q_2+2)$.}
%\begin{tabular}{c|c|c|c|c}
% $Q_2$ &  $14$ &
% $12$
% &  $8$ & 4 
%\\ \hline
% $\alpha(Q_2)$ &  $0.6$ &
% $0.6$
% & $1.74$ & 1.9 
%\end{tabular}
%\label{table2}
%\end{center}
%\end{table}

%\vspace{3.0cm}

We are left with clusters of charge $Q_2=10,6,2$ [case ($C$)]. 

They are eliminated successively by considering the jumps (\ref{jump}) in
the chemical potential for the cases in Table \ref{table_n1} (col. 5,6).
%\begin{table}[t]
%\begin{center}
%\caption{Successive elimination of charges $Q_2=10,6,2$ in the phase diagram
%(Fig. \ref{fig2}(B)) due to jumps (\ref{jump}) in the chemical potential.}
%\begin{tabular}{c|c|c}
% $(Q_1,Q_2,Q_3)$ & $\Delta(Q_2,\alpha)$ & $\alpha=\alpha(Q_2)$ \\ \hline
%(6,10,16) & $\frac{1}{12}\big[2E(16)+3E(6)-5E(10)\big]$ & $\alpha(10)=2.1\ldots$\\[5pt]\hline
%(2,6,16) & $\frac{1}{20}\big[2E(16)+5E(2)-7E(6)\big]$ & $\alpha(6)=2.2\ldots$\\[5pt]\hline
%(0,2,16) & $\frac{1}{14}\big[E(16)+7E(0)-8E(2)\big]$ & $\alpha(2)=3.5\ldots$\\[5pt]
%\end{tabular}
%\label{table4}
%\end{center}
%\end{table}
These jumps vanish at the couplings $\alpha=\alpha(Q_2)$ listed in the third 
column, such that the clusters with charges $Q_2$ can be eliminated for
$\alpha>\alpha(Q_2)$. The resulting phase diagram is shown in Fig. \ref{fig2}(B).
The pairs of integers in the rectangular domains denote the two cluster charges
which determine the ground state in the product ansatz.

Let us comment the appropriate boundary conditions for the
$L\times L$ cluster in the variational product ansatz for the ground
state on the infinite lattice.
A priori we are free in our choice of the clusters and their boundary
conditions. In our opinion periodic boundary conditions are most
appropriate for the following reasons:

The variational product ansatz becomes exact in the limit of infinite
cluster ($L\rightarrow\infty$, $N_c=L^2\rightarrow\infty$). In this
case, the cluster energies $E(Q,\alpha,N_c)=N_c\varepsilon(\rho_1=Q/N_c,\alpha)$
are related to the ground state energies per site $\varepsilon(\rho_1=Q/N_c,\alpha)$
at fixed charge density. On finite clusters with $L\times L=N_c$ sites
the ground state energies are approximated most accurately with periodic
boundary conditions, since the interaction between the clusters is partly
taken into account by means of the periodic boundary terms. Of course --
in a perturbation expansion for the interaction between the clusters --
the periodic boundary terms have to be subtracted again in each order. 
\section{Summary and discussion\label{sec6}}

In this paper we made an attempt to describe phase separation
in terms of a product cluster ansatz for the ground state of the
$2D$ $t-J$ model. 
%The clusters are of size $L\times L$ ($L=2,4$)
%and the ground state on these clusters are computed analytically
%(for $L=2$) and numerically (for $L=4$). 

The analytic results for $2\times 2$ plaquette clusters reveal
already the gross features as depicted in the phase diagram
in Fig. \ref{fig2}.
For $\rho$ and $\alpha$ large enough the ground state is built up
from two types of plaquettes $(Q_h,Q_e)=(0,4),(2,4)$. 
The hole clusters carry a charge $Q_h$ which increases with
decreasing $\alpha$:
%The first --
%called ``hole clusters'' -- carries the holes. Here the charge $Q_1$ is
\begin{eqnarray}
Q_h=0 \,\,\,\, &  & \hspace{5mm}\alpha(2)=4\sqrt{\frac{2}{3}}<\alpha 
\hspace{1.8cm}0\leq\rho\leq 1
 \label{regime_f1}\\
Q_h=2  \,\,\,\, &  & \hspace{5mm}\alpha(3)=\frac{2}{\sqrt{21}}\leq\alpha\leq \alpha(2)
\hspace{0.55cm}\frac{1}{2}\leq\rho\leq 1
\label{regime_f2}
\end{eqnarray}
where $\rho_1=Q_h/N$ denotes the charge density in the hole cluster.
The second type of plaquettes is completely occupied
with electrons $Q_e=4$.

In the regimes (\ref{regime_f1}) and (\ref{regime_f2}) the ground state
energy per site (\ref{ge_Q1Q2}) is linear in $\rho$ in the respective intervals.
%The interaction between neighbouring plaquettes -- as computed in first order
%perturbation theory in Section \ref{sec3} -- favours the condensation of
%plaquettes with maximal charge $Q_e=4$. This is the signature for
%phase separation.
The lower bounds in the $\alpha$-intervals listed in (\ref{regime_f1}), 
(\ref{regime_f2}) define two points on the boundary (\ref{large_rho})
for phase separation:
\begin{eqnarray}
\alpha_p(\rho_1=0)=4\sqrt{\frac{2}{3}}, \,\,\,\,\,\,& & \alpha_p(\rho_1=\frac{1}{2})=
\frac{2}{\sqrt{21}}\,.\label{R0}
\end{eqnarray}

The numerical results -- obtained from a product ansatz with $4\times 4$
clusters lead to the phase diagram in Fig. \ref{fig2}(B) similar to Fig. \ref{fig2}(A)
for $2\times 2$ clusters.
In each rectangular domain the cluster charges $Q_1,Q_2$ result from a
minimization of the ground state energy per site (\ref{ge_Q1Q2}) at fixed
charge density $\rho$ (in the infinite system). Phase separation is observed
in the upper part of Fig. \ref{fig2}(B)
with cluster charges $(Q_h,Q_e)=(0,16),(2,16),(6,16),(10,16)$
%, where the electron clusters
%have maximal charge $Q_e=16$. Holes occur in the clusters with
%charge $Q_1=Q_h<16$. 
The charge $Q_h$ of the hole clusters changes with the
spin exchange coupling $\alpha$
\begin{eqnarray}
Q_h=\,\,0 &  & \hspace{5mm}\alpha(2)=3.5<\alpha,\hspace{3.1cm} 0\leq\rho\leq 1
\hspace{0.4cm}\label{R1}\\
Q_h=\,\,2 &  & \hspace{5mm}\alpha(6)=2.2<\alpha<\alpha(2), \hspace{1.7cm}
\frac{1}{8}\leq\rho\leq 1\\
Q_h=\,\,6 &  & \hspace{5mm}\alpha(10)=2.1<\alpha<\alpha(6), \hspace{1.7cm}
\frac{3}{8}\leq\rho\leq 1\\
Q_h=10 &  & \hspace{5mm}\alpha(14)=0.6<\alpha<\alpha(10), \hspace{1.5cm}
\frac{5}{8}\leq\rho\leq 1\label{R4}
\end{eqnarray}

The ground state energy per site (\ref{ge_Q1Q2}) is linear in the $\rho$
intervals listed in (\ref{R1})-(\ref{R4}).
The lower bounds in the $\alpha$ intervals yield 4 points on the boundary
curve (\ref{large_rho}) for phase separation shown in Table \ref{table_n2}:

\begin{table}[ht!]
\begin{center}
\caption{Points on the boundary curve (\ref{large_rho}) for phase separation
derived from a $4\times 4$ cluster.}
\begin{tabular}{c|c|c|c|c}
%$\rho_1=\frac{Q_h}{16}$ & $0$ & $\frac{1}{8}$ & $\frac{3}{8}$ &
%$\frac{5}{8}$  \\ \hline
$\,\rho_1=Q_h/16\,$ & $\,0\,$ & $\,1/8\,$ & $\,3/8\,$ & $\,5/8\,$ \\ \hline
$\alpha_p(\rho_1)$ & 3.5 & 2.2 & 2.1 & 0.6
\end{tabular}
\label{table_n2}
\end{center}
\end{table}

\begin{figure}[ht!]
\centerline{\hspace{0.1cm}\includegraphics[width=5.0cm,angle=0]{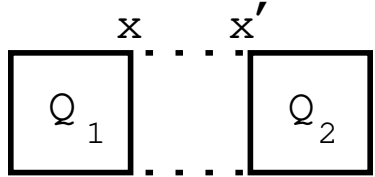}
\hspace{0.1cm}}
\caption{Interaction between neighbouring plaquettes with charges $Q_1,Q_2$
shown for a pair of nearest neighbour sites $x$ and $x'$.}
\label{fig3n}
\end{figure}

The product ansatz (cf. (\ref{prod_ansatz}) for plaquettes) with isolated
cluster ground states is highly degenerate. Each distribution of electron
and hole clusters over the whole lattice leads to the same ground state
energy, if we neglect the interaction between neighbouring clusters 
(cf. Fig.\ref{fig3n}).

%%%%%%%%%%%%%%%%%%%%%%%%%%%%%%%%%%%%%%%%%%%%%%%%%%%%%%%%%%%%%%%%%%%%%%%%
%%%%%%%%%%%%%%%%%%%%%%%%%%%%%%%%%%%%%%%%%%%%%%%%%%%%%%%%%%%%%%%%%%%%%%%%
Let us denote the interaction energy between neighbouring clusters with
charges $Q_2=L^2$, $Q_1=Q_h$ by $W(Q_1,Q_2)$ (cf. Fig.\ref{fig3n}). If the
difference
\begin{eqnarray}
\Delta & = & W(L^2,L^2)+W(Q_h,Q_h)-2W(Q_h,L^2)<0\label{ineq_Delta}
\end{eqnarray}
is negative, the ground state prefers phase separation in the following
sense. The numbers $N(L^2,L^2)$, $N(Q_h,Q_h)$ of identical neighbouring
clusters is maximal
\begin{eqnarray}
\frac{N(L^2,L^2)}{N}=2\frac{N(L^2)}{N} & ,\quad &
\frac{N(Q_h,Q_h)}{N}=2\frac{N(Q_h)}{N}\quad\quad\mbox{for\quad}N\rightarrow\infty
\label{Cfg_1}
\end{eqnarray}
where $N(L^2)$ and $N(Q_h)$ are the numbers of electron and hole clusters,
respectively. As a consequence the number of neighbouring clusters $N(Q_h,L^2)$
with different charges is minimal:
\begin{eqnarray}
\frac{N(Q_h,L^2)}{N} & \rightarrow & 0\quad\quad\mbox{for\quad}N\rightarrow\infty\,.
\label{Cfg_2}
\end{eqnarray}
Within the product ansatz (\ref{prod_ansatz}) with plaquettes of charge
$Q_2=L^2$, $Q_1=Q_h$ the interaction energies turn out to be
\begin{eqnarray}
W(Q_1,Q_2) =  \sum_x\langle Q_1,Q_2|h^{(x,x')}|Q_1,Q_2\rangle
 & = & -\frac{\alpha}{32}Q_1Q_2,  \;\;\;\; (L=2),
\label{relevant_ME}
\end{eqnarray}
provided that the cluster charges are even and the total cluster spins are zero.
In this case the spin matrix elements
\begin{eqnarray}
\langle Q_1|{\bf S}_x|Q_1\rangle=\langle Q_2|{\bf S}_{x´}|Q_2\rangle & = & 0
\end{eqnarray}
and the hopping contributions arising in (\ref{element_h_plaq}) vanish
for $L^2-Q_h\geq 2$. Eq. (\ref{relevant_ME}) implies that the inequality
(\ref{ineq_Delta}) is valid and the ground state configuration (\ref{Cfg_1}),
(\ref{Cfg_2}) with two big clusters of charge density $\rho_2=1$, $\rho_1=Q_h/L^2$
is selected out. According to (\ref{R1})-(\ref{R4}) the necessary condition
$L^2-Q_h\geq 2$ is satisfied for $\alpha>0.6$.

%%%%%%%%%%%%%%%%%%%%%%%%%%%%%%%%%%%%%%%%%%%%%%%%%
%abdollah-new

The phase diagrams in Fig. \ref{fig2}(A) and Fig. \ref{fig2}(B) contain
the whole information on the ground state of the system provided
that a product ansatz with two $L\times L$ clusters and charges
$Q_1$, $Q_2$ is adequate. Since we know the cluster ground states 
in each charge sector from the analytical
calculation in Ref. \cite{fl04} for $L=2$ and our numerical
calculation for $L=4$, we can also determine the hole--hole correlators:
\begin{eqnarray}
C_{(x,y)}(Q_1) & = & \langle n_h(0,0)n_h(x,y)\rangle -
\langle n_h(0,0)\rangle\langle n_h(x,y)\rangle
\label{h-h-c}
\end{eqnarray}
where $n_h(0,0)$, $n_h(x,y)$ count the number of holes at sites
$(0,0)$ and $(x,y)$. 

On the $4\times 4$ cluster with periodic boundary conditions
5 independent correlators can be arranged according to the distance vector ($x,y$).
%which are listed in Table \ref{table3}.
%\begin{table}[ht!]
%\begin{tabular}{c|c|c|c|c|c}
%$x-y$ & $(1,0)$ & $(1,1)$ & $(2,0)$ & $(2,1)$ & $(2,2)$
%\\ \hline
%correlator & $C_1$ & $C_{1,1}$ & $C_2$ & $C_{2,1}$ & $C_{2,2}$
%\end{tabular}
%\caption{
%}
%\label{table3}
%\end{table}
%Owing to the nearest neighbour couplings and the periodic boundary
%conditions, the $2D$ cluster $4\times 4=16$ is equivalent to the
%$4D$ cluster $2\times 2\times 2\times 2=16$ with an additional
%rotational symmetry which leads to the identity (Dagotto et al. \cite{dagotto90};
%Fabricius et al. \cite{fabricius91})
%\begin{eqnarray}
%C_{2,0} & = & C_{1,1}\,.
%\end{eqnarray}
%Let us start with the regime
Let us look at the following regime,
\begin{eqnarray}
2.2 & \leq & \alpha\leq 3.5,\hspace{2cm} \frac{1}{8}\leq\rho\leq 1\,. 
\end{eqnarray}
According to the phase diagram in Fig. \ref{fig2}(B) the ground state
contains clusters with charges $Q_1=2$, $Q_2=16$. The hole--hole
correlators (\ref{h-h-c}) in the $Q_1=2$ cluster turn out to be
negative.  However, the modulus of 
$|C_{1,0}(\alpha)|$ decreases with $\alpha$, whereas $|C_{2,0}(\alpha)|$, $|C_{2,1}(\alpha)|$
and $|C_{2,2}(\alpha)|$ increase with $\alpha$. 
We interprete
these increasing ``long''-range correlations as a hint to
condensation of holes for large $\alpha$-values.

Next let us look for the hole--hole correlators in the $\alpha$-regime
\begin{eqnarray}
0.6 & \leq & \alpha\leq 2.1,\hspace{2cm} \frac{5}{8}\leq\rho\leq 1\,. 
\end{eqnarray}
According to the phase diagram in Fig. \ref{fig2}(B) the ground state contains 
clusters with charges $Q_1=10$, $Q_2=16$. Again all correlators are negative.
The ``long''-range correlators $|C_{2,0}|=|C_{1,1}|$, $|C_{2,1}|$, $|C_{2,2}|$
are small in comparison with the nearest neighbour correlator $|C_{1,0}|$
for $0.6\leq\alpha\leq 1.5$. In this regime the system prefers formation
of hole pairs on neighbouring sites.

On the other hand
for $\alpha<0.6$ and $\rho>0.75$, we observe rapid changes of the ground
state with $\rho$ and $\alpha$ as is indicated by the numerous small
rectangular domains in the left upper part in Fig. \ref{fig2}(B). In this
regime holes are no longer confined as can be seen directly from the
product ansatz with $2\times 2$ plaquettes and charges $Q_h=3$, $Q_e=4$
[cf. upper left part in Fig. \ref{fig2}(A)]. Here, the hopping matrix elements
(\ref{element_h_plaq}) are active already 
in first order perturbation theory
and allow for the exchange of hole ($Q_h=3$) and electron ($Q_e=4$)
clusters. In principle, the holes can now hop over the whole lattice
and destroy thereby phase separation. The distribution of the holes
can be determined only from a precise computation of the ground state
in the low doping $\delta$, low $\alpha$ regime.

\end{document}